\documentclass[10pt, conference, compsocconf]{IEEEtran}

\ifCLASSINFOpdf
   \usepackage[pdftex]{graphicx}
   \usepackage{graphicx,dblfloatfix}

\else

\fi

\usepackage{color}
\usepackage{listings}
\begin{document}
%
\title{An Architecture for Decentralised Orchestration of Web Service Workflows}


\author{\IEEEauthorblockN{Ward Jaradat, Alan Dearle, and Adam Barker}
\IEEEauthorblockA{School of Computer Science, University of St Andrews, North Haugh, St Andrews, Fife, KY16 9SX, United Kingdom\\
\{ward.jaradat, alan.dearle, adam.barker\}@st-andrews.ac.uk}
}

\maketitle

\begin{abstract}

Service-oriented workflows are typically executed using a centralised orchestration approach that presents significant scalability challenges. These challenges include the consumption of network bandwidth, degradation of performance, and single-points of failure. We provide a decentralised orchestration architecture that attempts to address these challenges. Our architecture adopts a design model that permits the computation to be moved ``closer" to services in a workflow. This is achieved by partitioning workflows specified using our simple dataflow language into smaller fragments, which may be sent to remote locations for execution.

\end{abstract}


%
\IEEEpeerreviewmaketitle

\section{Introduction}
Service-oriented workflows are typically executed using a centralised orchestration approach. This approach allows workflows to be specified using an orchestration language, and executed on a central workflow engine. It provides process automation, and permits the workflow logic to be encapsulated and modified at a central location. However, this approach presents significant scalability challenges in high-performance and data-intensive workflows such as those seen in scientific applications [1]. The required input data to execute these workflows can be large in size, and it can increase gradually as more data becomes available from services during the workflow execution. This can cause a performance bottleneck as all the data pass through the central engine across the services in the workflow [2].\\

Decentralised orchestration is an alternative approach for executing service-oriented workflows. In this approach, distributed orchestration engines can collaborate together to execute the workflow, and each engine is responsible for executing part of the workflow specification. This permits intermediate data to be forwarded directly to the services that require it. Our research hypothesis states that this approach decreases the network traffic and improves the performance and execution time of workflows.\\

The main contributions of our work include a decentralised architecture that attempts to address the scalability challenges in service orchestration. Our architecture adopts a design model that permits the computation to be moved ``closer" to services in a workflow. It executes workflows based on our high-level dataflow specification language.

\section{Decentralised Orchestration Architecture}

Our architecture presents an approach for distributing web service workflow specifications to remote locations at which their execution takes place. This is achieved through partitioning a workflow specification into smaller fragments that may be transmitted to remote orchestration services for execution. The locus of control in centralised orchestration is represented by the central orchestration engine, which contains the decision logic for the workflow execution. However, the notion of a single locus of control does not exist in our architecture. The decision logic can be found at one or several orchestration services at any moment during the workflow execution.\\

\begin{figure}[h]
\centerline{\includegraphics[scale=0.8]{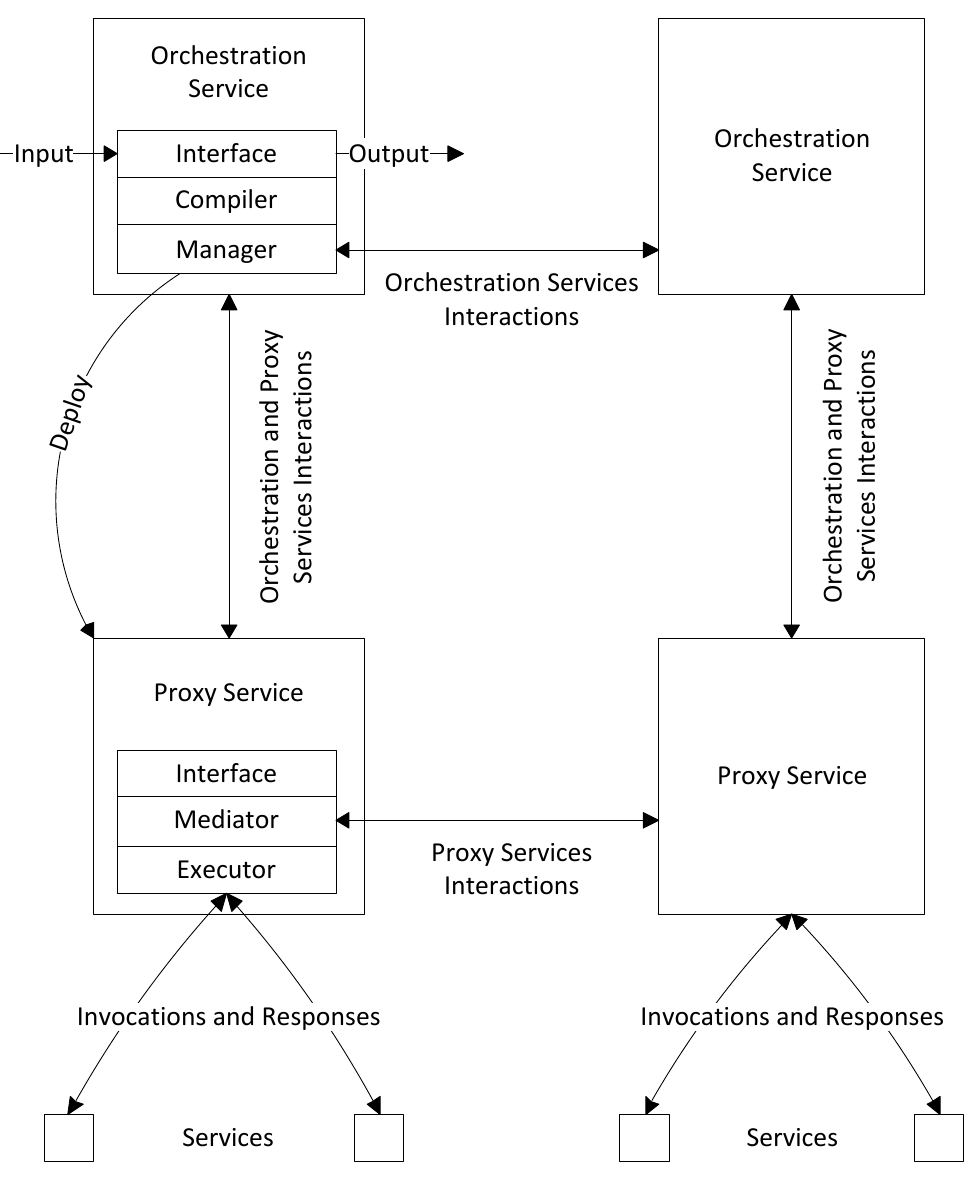}}
\label{fig:fig6}
\caption{Decentralised Orchestration Architecture}
\end{figure} 

\begin{figure*}[t]
\centerline{\includegraphics[scale=0.65]{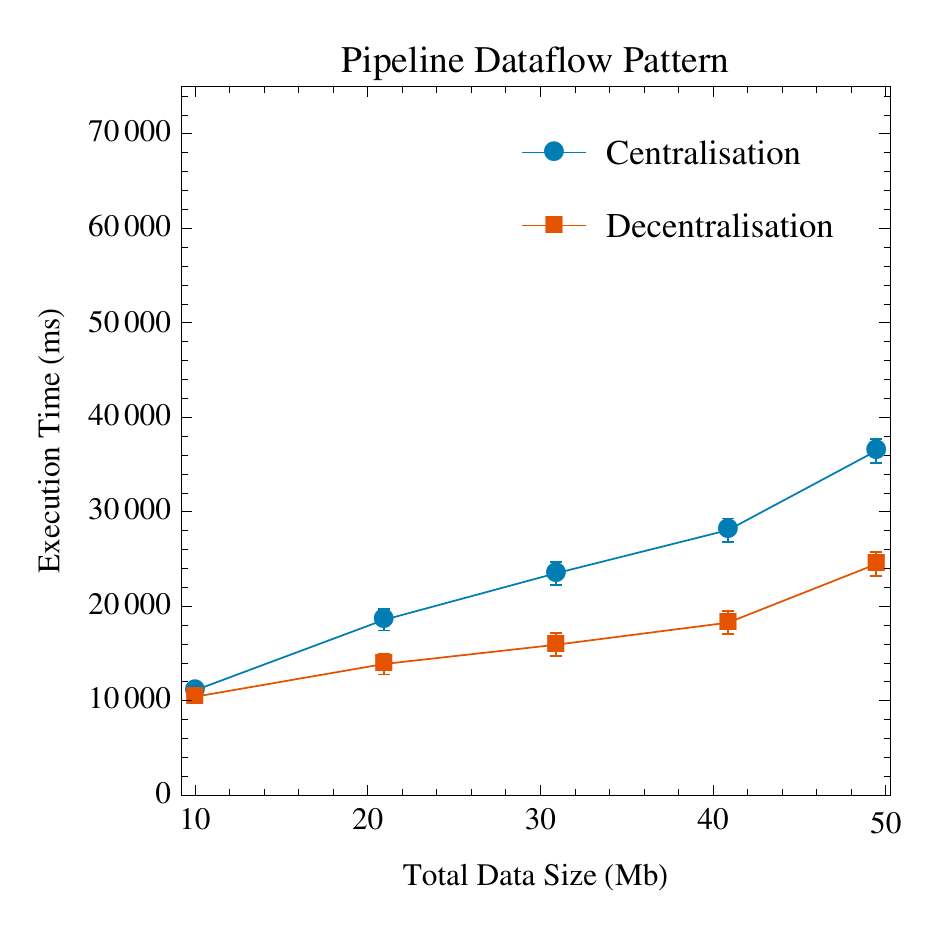} \includegraphics[scale=0.65]{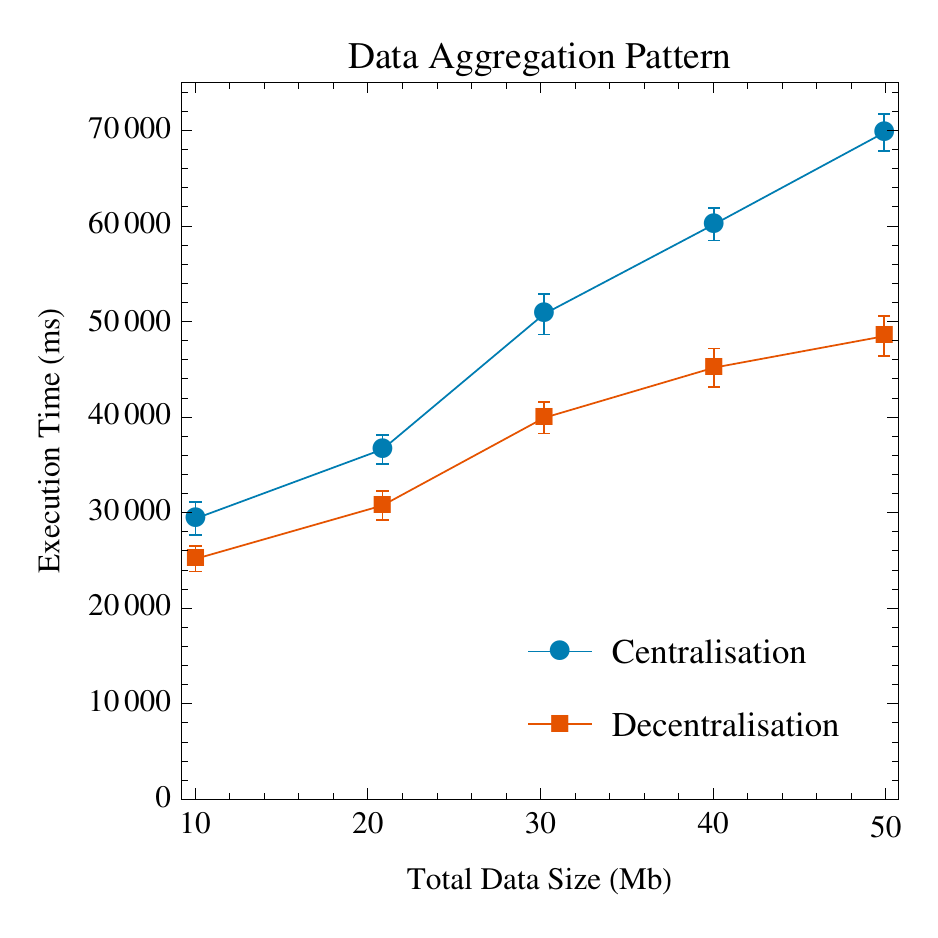} \includegraphics[scale=0.65]{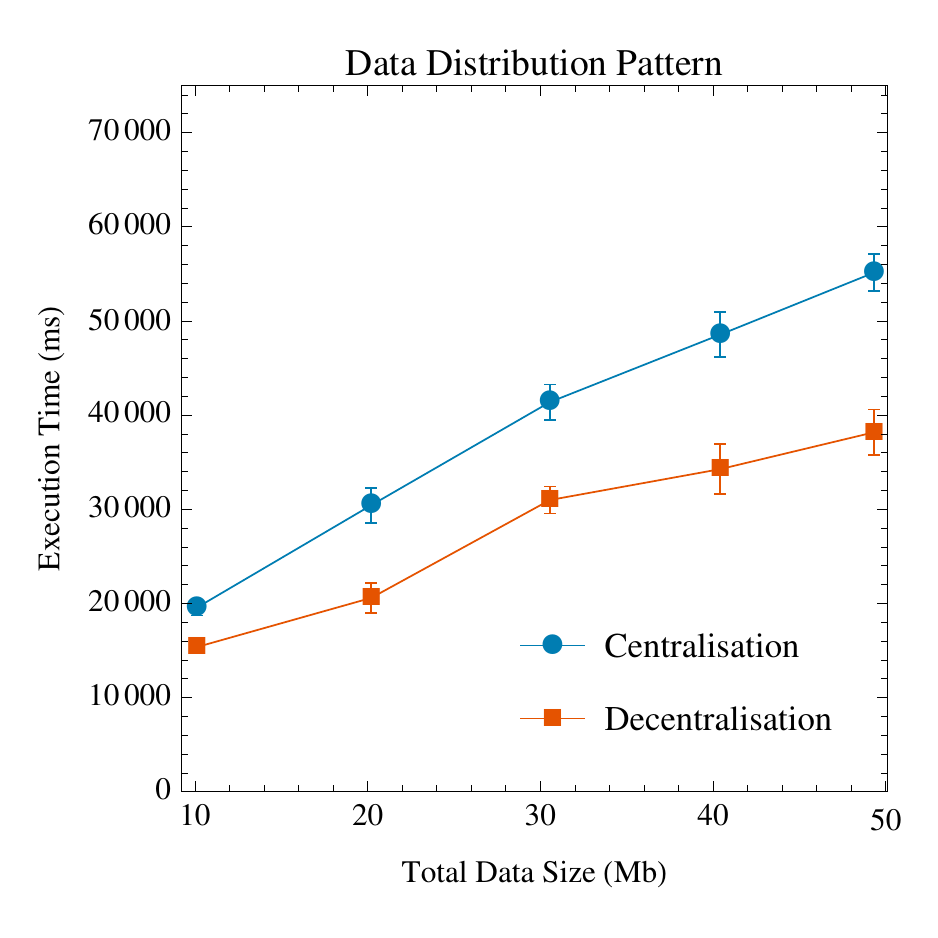}}
\label{fig:fig9}
\caption{Experimental Results}
\end{figure*}

Figure 1 provides an architecture overview diagram that shows the architecture components, and the interactions between them. These components consist of orchestration and proxy services. The orchestration service is responsible for analysing and partitioning the workflow specification into smaller fragments for execution at remote locations. It uses a compiler component to ensure the correctness of the workflow specification before executing it, and a manager component to interact with proxies. Proxy services are responsible for executing the workflow fragments. These proxies exploit connectivity to services in the workflow, and perform service invocations and compositions, data collection, retrieval, and mediation tasks on behalf of the orchestration services. Each proxy maintains state and data related to the workflow execution, and interacts with other proxies that may request its data to complete the execution of the overall workflow.\\

Our architecture executes web service workflows based on our dataflow specification language [3]. This language provides high-level abstractions that define a set of services and coordinate the dataflow between them. It separates between the workflow logic and its execution, supports implicit parallelism, and provides a data-driven execution model. It supports the specification of common dataflow patterns that can be combined together to create complex workflows, these include the pipeline, data aggregation and data distribution patterns [1].

\section{Evaluation}

We have conducted a set of experiments that aim to evaluate the performance of our architecture during the execution of workflows. In these experiments, web service workflows based on common dataflow patterns were executed by our architecture in centralised and decentralised configurations on Amazon EC2. The mean speedup rates for executing these workflows based on the pipeline, data aggregation, and data distribution patterns are 1.37, 1.30, and 1.41. For further information about these experiments please refer to [3].

Figure 2 displays a set of graphs that provide the total size of data communicated in each workflow, and the workflow execution time for each experiment. The performance analysis verifies our research hypothesis, and shows that our approach reduces the workflow execution time, and scales accordingly with the increasing size of data sets.

\section{Conclusion and Future Work}

This paper has presented our decentralised architecture that attempts to address the scalability challenges in service orchestration. In our approach, workflows based on our simple dataflow language can be executed by distributed orchestration services, which rely on proxies to exploit connectivity to web services in the workflow. Future work will include the investigation of workflow partitioning mechanisms, and providing execution policies to accommodate performance optimisation and resource utilisation requirements. It is our intention to address security requirements by providing information-flow policies to regulate the dissemination of confidential data in workflows. Further information about our work is available at http://bigdata.cs.st-andrews.ac.uk/.

\end{document}